\documentclass [12pt,fleqn]{article}

\usepackage{verbatim}
\usepackage{graphicx}
\usepackage{amssymb}
\usepackage{amsmath}
\usepackage{natbib}                 
\usepackage{enumerate}
\usepackage{tikz}
\usepackage{float}
\usepackage{url}

\usepackage[decisionutilitycolor]{influence-diagrams}
\NewDocumentCommand\voci{O{}D<>{1}m}{\incentive{#3}{incentive, gray,#1}{#2}}

\newtheorem {lemma} {Lemma}

\newtheorem {corollary} {Corollary}
\newtheorem {definition} {Definition}
\newtheorem {proposition} {Proposition}

\newcommand{\supp}{\operatorname{\emph{supp}}}
\newcommand{\rar}{\rightarrow}

\newcounter{excount}
\newenvironment{example}
{\refstepcounter{excount}\bigskip\noindent\textbf{Example~\arabic{excount}:}}{$\;\rule{1.5mm}{3mm}$\medskip}

\newcounter{asscount}

\newcommand{\eprf}{$\;\rule{1.5mm}{3mm}$ \smallskip}

\begin{document}

\title{Forecasting with Feedback\thanks{We thank Viola Grolmusz, Keri Hu, Navin Kartik, Malte Kn\"{u}ppel, Istv\'{a}n K\'{o}nya, Mats K\"{o}ster, Julian Martinez-Iriarte, Katrin Rabitsch, Joel Sobel, Ross Starr, Yuehui Wang, \'{A}d\'{a}m Zawadowski, an anonymous referee, and especially Darrel Cohen and Graham Elliott for comments and inspiration. Blandine Ledoux provided excellent research assistance. All errors are our responsibility. Robert Lieli gratefully acknowledges support by the National Bank of Austria's Jubil\"{a}umsfonds under project Nr.\ 19011.}}

\author{Robert P. Lieli\footnote{lielir@ceu.edu.}\\Department of Economics\\Central European University
\and Augusto Nieto-Barthaburu\footnote{augusto.nieto@gmail.com.}\\Department of Economics\\Universidad Nacional de Tucuman}
\maketitle

\thispagestyle{empty}

\begin{abstract}
\begin{footnotesize}
Systematically biased forecasts are typically interpreted as evidence of forecasters' irrationality and/or asymmetric loss. In this paper we propose an alternative explanation: when forecasts inform policy decisions, and the resulting actions affect the realisation of the forecast target itself, forecasts may be optimally biased even under quadratic loss. The result arises in environments in which the forecaster is uncertain about the policymaker's reaction to the forecast, which is presumably the case in most applications.  We motivate our theory by reviewing some stylised properties of Greenbook inflation forecasts. Our results point out that the presence of policy feedback poses a challenge to traditional tests of forecast rationality.


\bigskip



\end{footnotesize}	
\end{abstract}


\section{Introduction}

We model the production of point forecasts in an environment where a forecaster engages in strategic interaction with a decision maker (DM).\footnote{We will use the pronoun ``he'' to refer to the forecaster and ``she'' to the decision maker.} The reason why this interaction takes place is because the action taken by the DM in response to the forecast affects the realised value of the target variable that the forecast was meant to predict. Thus, in the language of our paper, forecast production is subject to \emph{feedback}. The following examples illustrate forecasting environments with and without feedback.

\begin{example}\label{ex: weather}
(No feedback) Many people use weather forecasts in their daily lives to make decisions that range from the trivial, e.g., whether to go hiking or to the movies, to the more consequential, e.g., whether to evacuate before a storm, or how much to irrigate a field. It is clear in this setting that the decisions made on the basis of the forecast do not affect the realised weather outcome. \eprf
\end{example}

\begin{example}\label{ex: cb}
(Feedback) Central bank staff inflation forecasts serve a dual role. First, as with any other forecast, they are meant to provide an accurate prediction of realised inflation over some horizon. Second, they also serve as guidance for setting the policy rate. Feedback occurs because interest rate decisions based on the forecast affect realised inflation in the future. \eprf
\end{example}

A simple way to model forecast production in the presence of feedback is to assume that the forecaster minimises expected loss \emph{conditional} on a pre-specified action available to the DM. Of course, the DM may end up deviating from the assumed action, naturally causing the forecast to be biased. Hence, conditional forecasts are not fully rational in that they are conditioned on potentially counterfactual actions. By contrast, we model the production of \emph{unconditional} forecasts, meaning that the forecaster anticipates and factors in the expected response of the DM to the forecast itself. The purpose of our paper is to study the statistical properties of such unconditional forecasts. 

Our model provides multiple insights into forecast production and evaluation in interactive environments. First, we show that in the presence of feedback \emph{and} some uncertainty about the DM's reaction, the optimal forecast is biased despite the forecaster possessing quadratic loss. Furthermore, the intercept and slope of the Mincer-Zarnowitz (MZ) regression function\footnote{The Mincer and Zarnowitz (1969) regression is the simple linear regression of the realised outcome on the forecast.} differ from zero and one, respectively, and the error associated with the optimal forecast is correlated with the forecaster's information. The mechanism that generates these results is a bias-variance tradeoff. When the forecast is used for policy, the forecaster recognises that the realised outcome depends on the forecast through the DM's reaction function. In case there is no uncertainty about this function, the forecaster can anticipate the DM's response perfectly, and, under quadratic loss, it is optimal to correct for it in full. Hence, the resulting forecast is unbiased. However, when the reaction function is subject to uncertainty, the volatility (variance) of the outcome also becomes dependent on the forecast. The forecaster can mitigate the volatility effect of his forecast by making it less sensitive to his information than the unbiased forecast would be. Doing so (up to a certain point) reduces the overall mean squared error, and hence the optimal forecast is biased.

Second, we provide equilibrium results where it is not only the forecaster who acts strategically (by attempting to anticipate the DM's response) but the DM herself holds correct beliefs about how the forecast is constructed and factors that into her reaction. We derive explicit formulas for the equilibrium bias and the MZ regression coefficients in terms of the mean and variance of the uncertain parameter in the DM's reaction function. The absolute value of the bias depends on the degree of uncertainty in the DM's reaction; lower uncertainty shrinks the bias toward zero. In addition, the MZ regression coefficients are shown to be nonlinear functions of the average strength and volatility of the DM's reaction. In particular, the MZ slope associated with the equilibrium forecast can be very flat or even negative, but, in line with the behaviour of the bias, it converges to one as uncertainty about the DM's reaction vanishes.

These results are clearly in violation of the canonical properties of optimal predictions under quadratic loss and no feedback. There is a long tradition in the forecasting literature of interpreting such violations as evidence of either ($i$) forecasters' ``irrationality'': \cite{mincer:1969aa}, \cite{zarnowitz:1985aa}, \cite{keane:1989aa, keane:1990aa}, \cite{romer:2000aa}, \cite{croushore:2012aa}, \cite{patton2012aa}, \cite{rossi:2016aa}, among many others; or ($ii$) asymmetries in the forecaster's loss function: \cite{granger:1969aa}, \cite{elliott:2005aa, elliott:2008ab}, \cite{patton:2007aa}, \cite{capistran:2008aa} and others. Our analysis shows that 
in the presence of feedback a third interpretation is possible, and that in this case the forecaster's loss function alone no longer determines the properties of optimal forecasts.

\paragraph{Empirical motivation}\label{sec: stylised facts}
Understanding why forecasts may be (optimally) biased has practical relevance. For instance, important economic forecasts exhibit systematic bias for reasons that are not immediately clear. As a case in point, we present some stylised properties of Greenbook (GB) inflation forecasts, produced by the Federal Reserve staff over several decades.\footnote{The Greenbook is now called the Tealbook, but we keep the traditional (and more widely known) name.} There is little question that feedback should be a relevant consideration in evaluating the properties of GB forecasts, since the Federal Open Market Committee uses these forecasts in formulating monetary policy, which then affects realised inflation in the future (among other economic variables).\footnote{GB forecasts are often viewed as conditional on a given policy (i.e., interest rate path), rather than strategically anticipating the FOMC's response. While the construction of GB forecasts does start with a conditioning assumption, this interest rate path is already chosen somewhat judiciously (\cite{reifscneider:2008aa}) and the final forecasts admittedly involve further judgmental adjustments (\cite{reifschneider:1997aa}). If these judgments are based on expectations of the FOMC's reaction to 
the forecast, the original conditioning assumption may lose its meaning, lending GB forecasts an unconditional flavor.}

We present two remarkable patterns that our theory can speak to:

\begin{enumerate}
    
    \item [(i)] GB inflation forecasts show systematic bias over extended periods, and the sign of the bias shifts over time.\footnote{We use the last release of the GDP deflator to compute realised inflation. However, the presented facts are robust to using the second release, which is also a popular choice in the literature.} 
    
    \item [(ii)] The statistical relationship between realised inflation and GB forecasts, as described by the MZ regression, changes quite radically from the mid-1970s to the late 2010s.
            
\end{enumerate}

\begin{figure}
    \centering
    \includegraphics[scale=0.7]{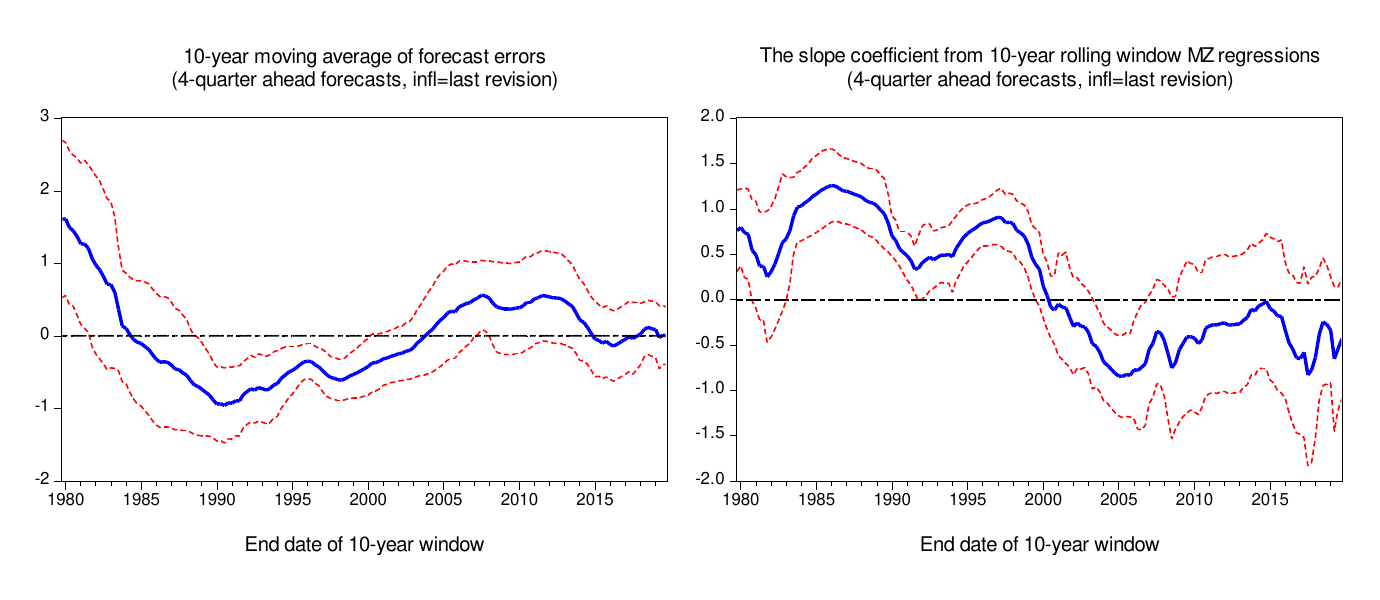}
    \caption{Properties of GB inflation forecasts: The evolution of the bias (left) and the MZ-slope (right). The dotted lines represent pointwise $\pm 2$ standard error bounds 
    (Newey-West standard errors with Bartlett kernel and bandwidth = 4 quarters). Data source: \cite{philly:2025aa, philly:2025bb}.
    }
    \label{fig:GB}
\end{figure}

The left panel of Figure~\ref{fig:GB} visually illustrates fact (i).
For a given quarter between 1980q1 and 2019q4, it shows the average error associated with the 4-quarter-ahead GB inflation forecasts over the preceding 40 quarters. Persistent but sign-changing bias is clearly a salient feature of GB forecasts, in line with the findings of other studies such as \cite{capistran:2008aa}.

Regarding point (ii), the right panel of Figure~\ref{fig:GB} depicts the slope coefficient from a rolling-window MZ regression of realised inflation on the same forecast over the same period. (The window length is again 40 quarters.) The evolution of the slope coefficient is nothing less than striking. It hovers around one from the mid-1980s to the early 1990s and, after a temporary drop, it again returns to unity by the end of the latter decade. However, the relationship between forecast and realisation shifts completely in the early 2000s. The slope coefficient dips into significantly negative territory by the middle of the 2000s, and then remains indistinguishable from zero in the last quarter or so of the sample period. While fact (ii) has not been presented in the literature in this form, it is consistent with some reported findings. For example, \cite{rossi:2016aa} conclude that: ``[GB] forecasts deteriorate over the 1990s and rationality tends to recover by the 2000s. However, for almost all forecast horizons, rationality breaks down again around 2005 (p.\ 526).'' Thus, the shifting nature of GB forecasts is confirmed by a formal test, but \emph{ibid.}\ interpret the result as a breakdown of rationality. Our paper points to another possible mechanism at play.

We do not claim that our model provides an exclusive or exhaustive explanation of the facts presented above. Inflation forecasting and monetary policy are complex processes, and there are many other factors that could contribute to the realised forecast error. The main goal of our paper is to highlight a specific mechanism---forecast-induced feedback---that yields patterns consistent with the data but has not been recognised in the literature as a driver of observed forecast properties. Moreover, the simple fact that forecast biases can arise for reasons other than irrationality or asymmetric preferences is an important caveat for the literature concerned with forecast rationality testing and loss function estimation. Studies engaged in these exercises either need to make the identifying assumption of no feedback explicit (when it is reasonable to do so), or think carefully about the feedback mechanism that could be present in the application.

The rest of the article proceeds as follows. In Section \ref{sec: litreview} we position our paper in several further threads of related literature. In Section \ref{sec: model} we present the model, and provide a thorough discussion of its underlying assumptions. In Section \ref{sec: results} we solve the forecaster's problem under feedback and study the statistical properties of optimal (and equilibrium) forecasts. Section \ref{sec: conclusion} concludes. Additional technical details of the model and the proofs of all formally stated results are contained in the Appendix.

\section{Related literature}\label{sec: litreview}

The dominant paradigm for modelling the production of point forecasts is to assume that the reported values result from minimising the expectation of a given loss function conditional on the forecaster's information; see, e.g., \cite{granger:1969aa}, \cite{stockman_1987aa}, \cite{christoffersen:1997aa}, \cite{elliott:2005aa}, \cite{patton:2007aa}, \cite{elliott:2008ab}, \cite{gneiting2011}. Through this exercise, the loss function ``picks out'' the statistical functional that maps the conditional predictive distribution of the outcome into an optimal point forecast, such as the mean functional in case of quadratic loss or the median functional in case of absolute loss. The theoretical properties of optimal forecasts follow from the properties of this functional. An alternative view, proposed by \cite{gneiting2011}, places the functional into a more primitive position; in particular, forecasters are assumed to work under a directive to report a given statistical functional associated with the predictive distribution. However, even in this case, a loss function consistent for the functional in question can be used to represent the forecast.\footnote{A small caveat is that some functionals, such as the mode, are not elicitable by any loss function; i.e., there is no loss function for which the mode is the optimal forecast.} Our paper adds to the forecasting theory literature by showing that in the presence of feedback the loss function is no longer sufficient to characterise optimal forecasts; the DM's reaction function is also a key determinant of forecast properties. 

The decision based perspective is also well established in the economic forecasting literature. Indeed, it is generally agreed that the main reason why government institutions and private entities engage in forecasting is to improve decision making; \cite{nelson:1964aa}, \cite{white1966aa}, \cite{granger:1969aa}, \cite{pesaran2004aa}, \cite{granger:2006aa}, \cite{patton:2007aa}, \cite{faust:2008aa} and a host of other authors have made this point at least in passing.\footnote{\cite{granger:2006aa} cite \cite{theil:1961aa} as an early systematic study of the links between forecasting and decision making.} In fact, the main theoretical argument for adopting an asymmetric loss function in the forecasting process is precisely because the economic cost of a decision made on the basis of a forecast that is too low is often different from the cost of a decision made on the basis of a forecast that is too high; see \cite{granger:1969aa} for classic examples. This would then imply that the optimal forecast is biased. Nevertheless, such decision based loss functions seem to be rarely employed in practice --- perhaps because it is not always easy to quantify losses or even identify a specific decision problem in which the forecast is to be used (see Example~\ref{ex: weather}). Thus, canonical loss functions, such as quadratic loss, remain the most commonly used. Our paper shows that optimal forecasts may be biased even under quadratic loss in the presence of feedback. 

The idea that forecasts might alter outcomes flows naturally from the decision based perspective, and has a long history in the social sciences. As an example of early work, \cite{hardt:2023aa} point to \cite{morgenstern:1928aa}, who speculated that forecast-induced feedback would necessarily invalidate public forecasts, making accurate prediction in many social settings impossible. This conjecture was later shown to be false by \cite{grunberg:1954aa} and \cite{simon:1954aa} in two studies that demonstrated the existence of a self-fulfilling forecast in different contexts.

Decades later, \cite{bowden:1987aa} used a sequential sampling model to study the convergence of forecasts and outcomes in situations where ``publication effects'' provide a causal link from the former to the latter. In a follow-up study, \cite{bowden:1989aa} foreshadows the use of game theory in modelling feedback, but does not give a formal treatment of this idea. These early frameworks are non-strategic in that the forecaster faces an exogenously given, deterministic reaction function. The formal results then centre on showing, via fixed-point arguments, that a self-fulfilling forecast exists. Thus, there is no room or need in these models to study the properties of forecasts. By contrast, in our model the equilibrium reaction function arises endogenously from the optimal behaviour of a strategic DM and is subject to uncertainty. This setup provides for qualitatively new results.

The interaction between forecasting and decision making has received some attention in the context of how central banks set policy (see Example~\ref{ex: cb}). In a model of professional inflation forecasts and monetary policy, \cite{bernanke:1997aa} study the existence of an informative rational expectations equilibrium (REE). They argue that it may not exist, but do not derive the properties of optimal forecasts in case there is an equilibrium. \cite{faust:2005aa} focus on comparing the information content for the public of conditional and unconditional central bank forecasts. On the empirical side, \cite{faust:2008aa} propose forecast efficiency tests for conditional forecasts, and \cite{knuppel:2017aa} compare inflation forecasts from various central banks with different assumptions about future monetary policy. But these papers do not give a theoretical account of the feedback mechanism and its consequences, which is our main focus. \cite{lieli:2020aa} deals with policy feedback in futures markets; there are analogies between forecasts and futures prices in that they can both be viewed as ``summaries'' of agents' private information. However, \textit{ibid.}\ focuses on studying the existence of a REE, while the focus here is on the statistical properties of forecasts. 

Finally, our theoretical model borrows its basic structure from communication games, pioneered by \cite{crawford:1982aa}. In particular, in our model the forecast is interpreted as a message, the forecaster plays the role of the ``sender'' of the message, and the DM the role of the ``receiver'' of the message. However, our setup departs from standard models of costless communication such as \cite{crawford:1982aa} in two ways. First, in our model the forecaster's message (i.e., the forecast) enters his loss function directly, making ours a game of ``costly talk,'' as in \cite{kartik:2007aa}. Second, we explicitly model the mechanism by which the state of the economy and the policy action jointly determine the realisation of the outcome. This causality is absent in standard communication games, where the realisation of the target variable is exogenous.

\section{The model}\label{sec: model}

We present the model in three steps. First, in Section \ref{subsec: forecasting environment} we describe the general forecasting environment. Next, in Section \ref{subsec: DM's problem} we model the DM's reaction to the forecast. That section also contains a visual summary of the model in the form of a timeline and an influence diagram. Finally, in Section \ref{subsec: discussion of assumptions} we discuss in detail the assumptions that we build into the first two steps. 

\subsection{The forecasting environment}\label{subsec: forecasting environment}

We consider the problem of an economic forecaster who, standing at time $t$, sets out to construct a forecast $f_t$ of an outcome $y_{t+1}\in \mathbb{R}$, where the value of the forecast may affect the realised outcome through an action prompted by it. Let $I_t$ denote the forecaster's information available at time $t$. The outcome $y_{t+1}$ is assumed to depend on $I_t$ through a scalar private signal $\theta_t=g(I_t)\in \mathbb{R}$, which we refer to as the \emph{state} of the economy, and which can be interpreted as a sufficient statistic for $I_t$. The DM does not directly observe $\theta_t$, and relies instead on the forecast to take an action $a_t\in \mathbb{R}$. We assume that $y_{t+1}$ depends on the state, the action, and a random error $\epsilon_{t+1}$ in a linear way:
\begin{eqnarray}\label{eqn: outcome equation}
y_{t+1}=\theta_t+ a_t + \epsilon_{t+1}.
\end{eqnarray}

The error $\epsilon_{t+1}$ is assumed to have zero mean and finite variance $\sigma^2>0$, and to be unforecastable in the sense of being independent of the forecaster's information (and hence $\theta_t)$. 
The state $\theta_t$ describes the evolution of the expected outcome in case the DM's action were held constant. For example, in the absence of policy changes, the outcome might follow an AR(2) process, in which case we can put $\theta_t=\phi_0+\phi_1y_t+\phi_2 y_{t-1}$. More generally, the mapping $\theta_t=g(I_t)$ may depend on many predictors contained in $I_t$ in a complicated way.\footnote{In the real world, it would be part of the forecaster's problem to model and estimate the function $g(\cdot)$. Here we take this function as a given.} 

The forecaster is endowed with a standard quadratic loss function
\[
L(y_{t+1},f_t)=(y_{t+1}-f_t)^2,
\]
and his goal is to construct a forecast that minimises expected loss (i.e., mean squared error, MSE) conditional on the observed value of $\theta_t$. Thus, if the forecaster expects the DM to use the reaction function $a_t=a(f_t)$, the forecaster's problem is to solve
\begin{equation}\label{eqn: fc prob general}
    \min_{f_t} E[(y_{t+1}-f_t)^2|\,\theta_t]=\min_{f_t} E\big\{[\theta_t+a(f_t)+\epsilon_{t+1}-f_t]^2\mid\theta_t\big\}.
\end{equation}
The solution of problem (\ref{eqn: fc prob general}) defines the optimal forecast $f^*_t=f^*(\theta_t)$ under the DM's assumed policy or reaction function.  

The fact that we allow the DM to choose her action based on the observed forecast of $y_{t+1}$ introduces the feedback effect that is the focus of our paper. Nevertheless, our framework also nests the traditional model of optimal forecasting with no feedback, where the outcome is assumed to be exogenous with respect to the forecast. Indeed, one can eliminate feedback in problem (\ref{eqn: fc prob general}) by setting $a(f_t)$ to a constant function, which can then be absorbed into the state $\theta_t$. It is well known that the MSE-optimal forecast is then given by the conditional mean, i.e., $f^*_t=E(y_{t+1}|\,\theta_t)=\theta_t$. Hence, the optimal forecast is unbiased and the associated forecast error is uncorrelated with the forecaster's information. 

It is not possible to study problem (\ref{eqn: fc prob general}) in detail without being more specific about the form of the reaction function $a(f_t)$ and where it comes from. This is the problem we address next.

\subsection{The decision maker's reaction function}\label{subsec: DM's problem}

\paragraph{The DM's objective} We equip the DM with a reaction function akin to a ``Taylor rule'' (\cite{taylor:1993aa}). More specifically, we assume that the DM has a target value $y^T$ for the outcome, which is common knowledge, and which she tries to achieve using the information provided by the forecast. Letting $E(\theta_t|f_t)$ denote the DM's expectation of $\theta_t$ conditional on the forecast, we assume that she chooses the action  
\begin{eqnarray}\label{eqn: R's optimal action}
a^*(f_t)=x_t\big[y^T-E(\theta_t|f_t)\big],
\end{eqnarray}
where $x_t$ is a positive multiplier.\footnote{This reaction function can be microfounded. In particular, Online Appendix~A shows that (\ref{eqn: R's optimal action}) can be derived as the solution to a simple control problem with a quadratic adjustment cost.}

Thus, the DM's objective is to close the gap between $y_{t+1}$ and $y^T$ that arises because the realised state may be different from the target value. Nevertheless, the DM's correction is generally imperfect. Indeed, for $x_t=1$, (\ref{eqn: R's optimal action}) implies $y_{t+1}=y^T$ up to the error $\theta_t-E(\theta_t|\,f_t)+\epsilon_{t+1}$, which is unforecastable on the basis of $f_t$. However, for $x_t\in(0,1)$ the action taken is insufficient to counteract the impact of the expected state on $y_{t+1}$ (e.g., because of adjustment costs), while for $x_t>1$ the DM overreacts, resulting in a gap with the opposite sign (e.g., because of political reasons or because the DM may care about other outcomes as well). We henceforth assume that the value of $x_t$ is drawn by ``nature'' from a distribution, independently of $(I_t,\epsilon_{t+1})$ and hence of $(\theta_t,\epsilon_{t+1})$. Thus, importantly, the value of $x_t$ is \emph{not} contained in the forecaster's information set $I_t$; it is, however, observed by the DM, i.e., it is the DM's private information. The forecaster is only assumed to know the distribution of $x_t$. This uncertainty about the DM's reaction to the forecast is a key driver of our results. 

\paragraph{The DM's expectations} In order to implement the policy rule (\ref{eqn: R's optimal action}), the DM must make a conjecture about the form of the functional relationship between $\theta_t$ and $f_t$, and use it to compute the expectation $E(\theta_t|f_t)$. Suppose that the DM conjectures that the forecast is linear in $\theta_t$; specifically, let $f_t=b+c\cdot \theta_t$, $c\neq 0$. Then the conditional expectation in (\ref{eqn: R's optimal action}) can be computed as
\begin{eqnarray*}\label{eqn: DM's expectations}
E(\theta_t|f_t)=\frac{f_t-b}{c}.
\end{eqnarray*}
Given the way in which the DM forms expectations, her optimal action (\ref{eqn: R's optimal action}) becomes
\begin{align}\label{eqn: R's optimal action with exp}
a^*(f_t)=x_t\left(y^T-\frac{f_t}{c}+\frac{b}{c}\right).
\end{align}
We will shortly see that the assumption of a linear forecasting rule is self-confirming; that is, when the DM conjectures a linear forecast function and uses it for policy decisions, it is indeed optimal for the forecaster to use a linear forecasting rule.

At this point we do not require that the intercept and slope conjectured by the DM be equal to the actual intercept and slope used by the forecaster. Instead, in the general case we allow for the DM to be mistaken about the forecaster's strategy. This generality helps accommodate applications in which policy follows a rigid, mechanical rule. The special case in which the DM's conjecture is correct constitutes an \emph{equilibrium} in our model, which we formally define in Section \ref{sec: results}. Furthermore, the assumption that the DM conjectures a linear forecast implies that we will be looking for \emph{linear} equilibria. We justify this focus in the next subsection.

\begin{example}\label{ex: Taylor}
If the DM conjectures $c=1$ and $b=0$, the reaction function (\ref{eqn: R's optimal action with exp}) reduces to the particularly simple Taylor rule
\[
a^*(f_t)=x_t(y^T-f_t).
\]
An interpretation that rationalises this reaction function is that the DM takes the forecast at ``face value'' and considers it, somewhat naively, equal to the state or, equivalently, the expected outcome given no action ($a_t=0$). Alternatively, this may be a behavioural rule without explicit justification. Either way, as we will show, the optimal forecast under this reaction function is not $f(\theta_t)=\theta_t$. Therefore, the DM's conjecture would not be part of an equilibrium.\eprf
\end{example}

Figure \ref{fig: infl diag} provides a visual summary of the model in the form of a timeline and an influence diagram.

\begin{figure}[!h]
\begin{center}
\begin{tikzpicture}

    \draw (0,0) -- (12,0);
   
    \foreach \x in {2,6,10}
    \draw (\x cm,3pt) -- (\x cm,-3pt);

   \draw (2,0) node[above=3pt] {$t$} node[below=3pt] {forecaster} node[below=15pt] {observes state $\theta$} node[below=27pt] {and produces} node[below=39pt] {forecast $f(\theta)$};
   \draw (6,0) node[below=3pt] {DM observes} node[below=15pt] {$x$ and $f$ and} node[below=27pt] {chooses} node[below=39pt]{action $a^*(f)$};
   \draw (10,0) node[above=3pt] {$t+1$} node[below=3pt] {Outcome $y$} node[below=15pt] {and payoffs} node[below=27pt] { realise};
  \end{tikzpicture}

\bigskip

\begin{influence-diagram}
  \node (U1) [utility] {Forecaster's \\ loss};
  \node (help1) [above = of U1, draw=none] {};
  \node (help2) [right = of U1, draw=none] {};
  \node (help3) [left = of U1, draw=none] {};
  \node (help4) [below = of U1, draw=none] {};
  \node (S) [above = of help3] {State ($\theta$)};
  \node (P1) [above = of help2, decision, player1] {Forecast ($f$)};
  \node (P2) [right = of help4, decision, player2] {Policy ($a$)};
  \node (O) [below = of help3] {Outcome ($y$)};
  \node (U2) [below = of help4, utility] {DM's loss};
     
  
 \edge {S} {P1};
 \edge {S} {O};
 \edge {P1} {P2};
 \edge {P1,O} {U1};
 \edge {P2} {O};
 \edge {P2,O} {U2};
  \edge {P1} {U1};

\end{influence-diagram}
\end{center}
\caption{The model's timeline and influence diagram. In the influence diagram uncertainty nodes are in white, value nodes in yellow, the forecaster's decision node is in red, and the DM's decision node is in purple.}
\label{fig: infl diag}
\end{figure}

\subsection{Discussion of the modelling assumptions}\label{subsec: discussion of assumptions}

\paragraph{Outcome determination} A linear outcome equation is a simplification that allows us to make our points in a clear and tractable way. Linearity, coupled with the independence of $\theta_t$ and $\epsilon_{t+1}$, imply that the DM's action has a direct impact only on the conditional mean $E(y_{t+1}|\,\theta_t)$ and not on higher moments of the outcome. Nonetheless, the feedback mechanism described in our model will be present qualitatively even if policy affects the distribution of the outcome in a more complex way.

\paragraph{Forecaster's preferences} Endowing the forecaster with quadratic loss is a deliberate choice.  As we will see in Section~\ref{sec: results}, the optimal forecast under feedback is generally biased. Using squared-error loss ensures that this result is not due to a ``built-in'' preference for under- or overprediction.   

\paragraph{Forecaster's information} Two key assumptions of the model are that $(i)$ the state $\theta_t$ is the forecaster's private information, and $(ii)$ the state cannot be communicated to the DM directly, only through the forecast. These assumptions are meant to capture, in an abstract way, how DMs and forecasters typically interact in practice. First, DMs often rely on advisors because they may lack the capacity or resources to acquire and process all information relevant for their decisions. Second, for similar reasons, the advice given by experts often comes in the form of summary statistics and projections of key outcomes rather than raw data. Indeed, professional forecasting would be unnecessary in the real world if constraints justifying $(ii)$ did not exist.

\paragraph{Linear forecast} In the model, the DM conjectures that the forecast is a linear function of the state. We then look for linear equilibria where this conjecture is self-confirming. Focusing on linear equilibria is common in economic theory because they are appealing as behavioural predictions and are analytically tractable. Indeed, we will explicitly solve our model for such an equilibrium and study it directly. 

\paragraph{Interpretation of $x_t$} The parameter $x_t$ in the DM's reaction function determines the strength of the policy action taken in response to the reported forecast. It also embodies any and all of the uncertainty that the forecaster faces with respect to policy. There may be several reasons in practice why the forecaster does not perfectly foresee the DM's planned action. First, the DM may be a committee, and the committee members may not fully agree on the appropriate action. For example, the FOMC publishes their projections of the policy rate as a ``dot plot,'' as they typically vary from member to member. Hence, the forecasting staff may justifiably be uncertain about which policy will prevail. Second, even when the DM is a single person, she may not be willing to communicate her precise intentions to the forecaster, as they may be part of a confidential plan. For instance, a firm's manager may not be willing to share their detailed strategy with the forecasters in order to prevent these plans from leaking to the competition. Third, the actions of the DM (or the effectiveness of those actions) may be subject to inherent randomness. This is manifested, for instance, by the practice of adding exogenous monetary policy shocks to macroeconomic models (and to the Fed's reaction function in particular). 

\paragraph{Independence of $x_t$ and $\theta_t$} We assume that the strength of the policy reaction  is independent of economic conditions. More generally, it is conceivable that the mean or variance of $x_t$ could depend on $\theta_t$; for example, a central bank may react more aggressively during a financial crisis, or there may be more uncertainty about the central bank's reaction in such times. However, there are many ways to model the possible dependence between $x_t$ and $\theta_t$, and it is not obvious what salient facts (if any) should be captured by this additional feature. In an abstract model such as ours independence is a natural benchmark.

\paragraph{Forecaster's expectations about policy} As discussed above, we assume that the forecaster knows the DM's reaction function $(\ref{eqn: R's optimal action with exp})$ up to the uncertain parameter $x_t$. It may seem contradictory that $x_t$ is the DM's private information while her conjecture about $(b,c)$ is not. This modelling choice can be rationalised in several ways. First, some forecasters, such as those in a central bank, may interact with the DM on a regular basis, and know a lot, but not everything, about her reaction function. We represent this residual uncertainty about policy via $x_t$ rather than $(b,c)$. Second, $(b,c)$ might be common knowledge if the DM operates under set rules or institutional constraints that call for a certain interpretation of the forecast (see Example~\ref{ex: Taylor}). Third, the assumption is certainly true in equilibrium, where by definition the DM's conjecture about $(b,c)$ coincides with the values actually used by the forecaster. Finally, forecasts derived under an arbitraty $(b,c)$ vector can be regarded simply as technical objects which help derive the equilibrium forecast in the model.

\paragraph{Forecaster's inability to commit} In our model the forecaster lacks the power to commit to a fixed forecasting rule ex-ante, which is natural given that the state is not directly observable by the DM. Modelling forecasters with commitment is a straightforward extension of our setup; commitment can actually increase forecaster welfare in equilibrium.

\section{Analysis of the model}\label{sec: results}

\subsection{The bias-variance tradeoff faced by the forecaster}\label{subsec: mean-variance tradeoff}

The forecaster's MSE objective (\ref{eqn: fc prob general}) admits a standard decomposition as the sum of the conditional volatility of the outcome and the square of the conditional bias of the forecast: 
\begin{eqnarray}
E[(y_{t+1}-f_t)^2|\theta_t] & =  & E\Big\{\big[y_{t+1}-E(y_{t+1}|\theta_t)\big]^2\big|\,\theta_t\Big\}+[E(y_{t+1}|\theta_t)-f_t]^2\nonumber\\
&=&Var(y_{t+1}|\theta_t) + bias^2(f_t|\theta_t)\nonumber\\
& = & Var[a^*(f_t)|\theta_t] + \big\{\theta_t+E[a^*(f_t)|\theta_t]-f_t\big\}^2 +\sigma^2,\label{eq: mse decomposition}
\end{eqnarray}
where the last equality uses the outcome equation (\ref{eqn: outcome equation}) and the independence of $\theta_t$, $x_t$, and $\epsilon_{t+1}$.\footnote{See the proof of Proposition~\ref{prop: non-strat opt forecast} in Online Appendix~C for a formal derivation.}  

In the absence of uncertainty about the policy response, $Var[a^*(f_t)|\,\theta_t]=0$, so the first term in (\ref{eq: mse decomposition}) vanishes and the optimal forecast simply drives the bias term to zero. As in this case $E[a^*(f_t)|\,\theta_t]=a^*(f_t)$, this is accomplished by solving the fixed point problem  $\theta_t+a^*(f_t)=f_t$ for $f_t$. The interpretation is that when the DM's response to the forecast is perfectly foreseeable, the optimal forecast will correct for it in full. By contrast, if the policy response is uncertain then $Var[a^*(f_t)|\,\theta_t]$ is generally a function of the forecast $f_t$. That is, the choice of the forecast contributes to the overall MSE not only through its systematic error but also through its impact on the volatility of the outcome. This gives rise to a non-trivial bias-variance tradeoff in (\ref{eq: mse decomposition}); the optimal resolution of this tradeoff calls for manipulating the DM's action through the forecast in a way that causes the optimal forecast to be biased, as we show below.

To derive more concrete results, we now make use of the specific form of the reaction function in equation (\ref{eqn: R's optimal action with exp}), which incorporates the DM's interpretation of the forecast as a linear forecasting rule $f_t=b+c\cdot \theta_t$. Denoting the mean and variance of the ``strength of policy'' parameter $x_t$ by
\[
\mu:=E(x_t)\;\text{ and }\;\tau^2:=Var(x_t),
\]
respectively, we obtain
\begin{eqnarray}
    &&Var(y_{t+1}|\,\theta_t)=Var\big[a^*(f_t)\big|\,\theta_t \big]+\sigma^2=\tau^2 \left(y^T-\frac{f_t}{c}+\frac{b}{c}\right)^2 + \sigma^2 \label{eqn: var expansion}
\end{eqnarray}
and
\begin{eqnarray}
    &&bias^2(f_t|\,\theta)=\big\{\theta_t+E[a^*(f_t)|\theta_t]-f_t\big\}^2 =\left\{\theta_t+ \mu \Big(y^T+\frac{b}{c}\Big)-\Big(\frac{\mu+c}{c}\Big)f_t\right\}^2 \label{eqn: bias expansion};
\end{eqnarray}
see the proof of Proposition \ref{prop: non-strat opt forecast} in Online Appendix~C for a more detailed derivation of these formulas. The forecaster's problem (\ref{eqn: fc prob general}) is equivalent to minimising the sum of expressions (\ref{eqn: var expansion}) and (\ref{eqn: bias expansion}) with respect to $f_t$.

The bias-variance tradeoff that the forecaster faces can be seen by considering equations (\ref{eqn: var expansion}) and (\ref{eqn: bias expansion}). The forecaster could choose a forecasting rule that reduces bias to zero; in particular, from (\ref{eqn: bias expansion}) we see that the unbiased forecast is indeed a linear function of $\theta_t$ with a slope of
\begin{eqnarray*}
\frac{c}{\mu+c}
\end{eqnarray*}
and an appropriate intercept term. However, the optimal forecast is different from the unbiased forecast because the variance of the outcome, i.e., equation (\ref{eqn: var expansion}), also depends on the forecast. 
In particular, the optimal forecast consists of a linear forecasting rule with slope
\[
\lambda\cdot \frac{c}{\mu+c},
\]
where $\lambda=(\mu+c)^2/[\tau^2+(\mu+c)^2]\in (0,1)$; see Proposition \ref{prop: non-strat opt forecast} below. Hence, we see that the forecaster tries to reduce the volatility of the outcome by attenuating the slope toward zero relative to the unbiased forecast. That is, he diminishes the sensitivity of the forecast to the state $\theta_t$, in an attempt to mitigate the volatility of the policy response. The cost of this manipulation is the introduction of a systematic error.  

We summarise and slightly extend the preceding discussion via a formal proposition.
\begin{proposition}\label{prop: non-strat opt forecast}
Given the DM's reaction function (\ref{eqn: R's optimal action with exp}), 
the optimal forecast is of the form $f^*_t = d^*+e^* \cdot \theta_t$,
where 
\begin{eqnarray}\label{eqn: non-strat intercept}
d^*=c \left[\frac{\tau^2+\mu(\mu+c)}{\tau^2+(\mu+c)^2}\right] \left(y^T+\frac{b}{c}\right) ,
\end{eqnarray}
and
\begin{eqnarray}\label{eqn: non-strat slope}
e^*=c \frac{\mu+c}{\tau^2+(\mu+c)^2}.
\end{eqnarray}
\end{proposition}

Proposition~\ref{prop: non-strat opt forecast} shows that the DM's conjecture of a linear forecasting function is self-confirming, i.e., that given such a conjecture, the optimal forecast $f^*_t=f^*(\theta_t)$ is indeed a linear function of $\theta_t$. Nevertheless, the DM's guesses about the slope and the intercept of $f^*(\theta_t)$ that are built into the reaction function are not required to be correct at this point. The following example illustrates the optimal forecasting rule against the simple Taylor rule of Example~\ref{ex: Taylor}.

\begin{example}\label{ex: Taylor opt forecast}
If $c=1$ and $b=0$, i.e., $a^*(f_t)=x_t(y^T-f_t)$, then Proposition~\ref{prop: non-strat opt forecast} gives $f^*_t = d^*+e^* \cdot \theta_t$ with
\[
d^*=\frac{\tau^2+\mu(\mu+1)}{\tau^2+(\mu+1)^2}y^T\;\text{ and } \;e^*=\frac{\mu+1}{\tau^2+(\mu+1)^2}.
\]
This confirms that the optimal forecast under the naive Taylor rule is not $f(\theta_t)=\theta_t$; instead, as explained above, the slope is attenuated toward zero, and the intercept is proportional to the target $y^T$. Thus, the DM's interpretation of the forecast is not consistent with how it is actually produced.\eprf
\end{example}

\subsection{The equilibrium forecast}
We now consider forecasting in the presence of a fully strategic DM, whose reaction function correctly anticipates the forecasting rule used. We formally define a (linear) equilibrium as follows. 

\begin{definition}\label{definition: equilibrium}
A linear equilibrium forecast is a function $f^\dag (\theta)=b^\dag+c^\dag \cdot \theta$ such that if the DM conjectures the use of $f^\dag(\theta)$, then it is optimal for the forecaster to use $f^\dag(\theta)$. More formally, if the DM sets $b=b^\dag$ and $c=c^\dag$ in (\ref{eqn: R's optimal action with exp}), then $d^*=b^\dag$ and $e^*=c^\dag$ in Proposition~\ref{prop: non-strat opt forecast}. 
\end{definition}

The equilibrium concept that underlies Definition \ref{definition: equilibrium} is pure-strategy Perfect Bayesian Equilibrium (PBE).
In equilibrium the forecaster and the DM mutually best-respond to each other, and hold correct (i.e., rational) expectations. The following corollary to Proposition~\ref{prop: non-strat opt forecast} shows that a linear PBE exists in our model if the uncertainty about the DM's reaction is below a threshold.
\begin{corollary}\label{corollary: equilibrium slopes}
For $\tau^2\le \frac{1}{4}$ there exists a linear PBE with equilibrium forecast $f^\dag(\theta_t)=b^\dag+c^\dag\cdot\theta_t$, where 
\begin{align}\label{eqn: equilibrium slope}
c^\dag =\frac{1}{2}-\mu+\frac{\sqrt{1-4\cdot \tau^2}}{2}
\end{align}
provided that $c^\dag\neq 0$, and
\begin{align}\label{eqn: equilibrium intercept}
b^\dag =(1-c^\dag) y^T.
\end{align}

\end{corollary}

\paragraph{Remarks} 
\begin{enumerate}

\item The equilibrium forecast slope is derived from equation (\ref{eqn: non-strat slope}) by substituting the common value $c^\dag$ for $c$ and $e^*$ and then solving for it. The equilibrium intercept is then obtained analogously from equation (\ref{eqn: non-strat intercept}); see Online Appendix~C for details.

\item The equilibrium forecast is a one-to-one function of the state; hence, a rational DM can learn the true value of the state from the forecast. That is, the equilibrium is fully revealing (or ``separating'' in the jargon of game theory).

\item If $\tau^2>\frac{1}{4}$ a linear equilibrium does not exist. Indeed, it is easy to verify that if $\tau^2>\frac{1}{4}$, then $\frac{\mu+c}{\tau^2+(\mu+c)^2}<1$ for all values of $c$ and $\mu$. Hence, equation (\ref{eqn: non-strat slope}) implies that for any slope $c$ conjectured by the DM, the forecaster will always want to use a slope that is attenuated toward zero compared with the DM's conjecture. 
If $\supp(x_t)\subseteq [0,1]$, the condition $\tau^2\le \frac{1}{4}$ is always satisfied, i.e., in this case an equilibrium always exists. 

\item The equilibrium forecast slope $c^\dag$ can be negative for some parameter configurations (large $\mu$, large $\tau^2$). This may seem odd given that a higher value of $\theta_t$ leads to a higher value of $y_{t+1}$ \emph{ceteris paribus} (see equation (\ref{eqn: outcome equation})). Nevertheless, $c^\dag$ also incorporates the DM's intervention to the forecast, and this intervention goes against (and may overcompensate for) the higher state. Furthermore, $c^\dag>0 \implies b^\dag<y^T$ and $c^\dag<0 \implies b^\dag>y^T$, while $sign(b^\dag)=sign(y^T)$ because $c^\dag\le 1$.

\item In solving for the equilibrium in Corollary \ref{corollary: equilibrium slopes}, it is immediately seen that there is an alternative equilibrium with a different forecast slope. This equilibrium is presented in Online Appendix~B along with arguments for why we choose to focus on the equilibrium in Corollary~\ref{corollary: equilibrium slopes}.

\end{enumerate}

\subsection{The statistical properties of optimal forecasts under feedback}\label{subsection: statistical properties}

The following proposition represents the central result of the paper: an explicit  characterisation of the conditional bias and the MZ regression coefficients associated with the optimal forecast presented in Proposition~\ref{prop: non-strat opt forecast}. The properties of the equilibrium forecast follow as a corollary. 

\begin{proposition}\label{proposition: statistical properties}
Let $f^*_t=f^*(\theta_t)$ denote the optimal forecast described in Proposition~\ref{prop: non-strat opt forecast}. Then:

\begin{enumerate}
\item [(a)] The optimal forecast is biased, and the bias conditional on the state is given by
\begin{align}\label{eq: bias}
E(y_{t+1}-f^*_t|\,\theta_t)=\frac{\tau^2}{\tau^2+(\mu+c)^2} (\theta_t-cy^T-b).
\end{align}

\item [(b)] The Mincer-Zarnowitz regression function of the optimal forecast is given by
\begin{align}\label{eq: MZ}
E(y_{t+1}|f^*_t)=-\frac{\tau^2}{\mu+c}\left(y^T+\frac{b}{c}\right)+\frac{\tau^2+c(\mu+c)}{c(\mu+c)}f^*_t.
\end{align}
\end{enumerate}
\end{proposition}

Proposition \ref{proposition: statistical properties} shows that optimal forecasts in our feedback model are generally biased (despite the forecaster's quadratic loss), and the slope of the MZ regression is different from one, implying that the forecast error $y_{t+1}-f^*_t$ is correlated with $f^*_t$. These properties continue to hold even when the DM is fully strategic (see Corollary \ref{corollary: equil bias MZ} below). As we pointed out in Section~\ref{subsec: mean-variance tradeoff}, the driving force behind these results is a bias-variance tradeoff induced by the combination of two factors: $(i)$ the \emph{feedback effect from forecast to realisation} and $(ii)$ the forecaster's \emph{uncertainty about the strength of the policy reaction}, i.e., $x_t$. Both mechanisms are necessary to induce bias in the optimal forecast. 

To assess the role of uncertainty, one can take the limit $\tau^2\to 0$ in equations (\ref{eq: bias}) and (\ref{eq: MZ}). It is immediately seen that even in the presence of feedback, when uncertainty about $x_t$ vanishes, bias vanishes, the slope of the MZ regression function converges to 1, and the intercept converges to zero. This confirms the general insight from Section~\ref{subsec: mean-variance tradeoff} that feedback from a deterministic reaction function does not cause bias because it is possible to anticipate it perfectly and then correct for it.\footnote{Even when uncertainty about policy is absent, the forecast with feedback will be a different function of $\theta_t$ than the forecast with no feedback (i.e., when $a(f)$ is a constant function).}

To see the effect of policy uncertainty without feedback,  suppose that the DM does not rely on the forecast to take action. In this case, the DM's optimal action is $a^*_t=x_t [y^T-E(\theta_t)]$; that is, she chooses an adjustment based on her unconditional expectation of the state. In turn, the forecaster's problem is 
\begin{align*}
\min_{f_t} E[ (y_{t+1}-f_t)^2 |\, \theta_t]=\min_{f_t} E\big\{[\theta_t + x_t(y^T-E(\theta_t))+\epsilon_{t+1}-f_t]^2 \big|\, \theta_t\big\}.
\end{align*}
This is then a perfectly standard prediction problem whose solution is the conditional mean $f^*(\theta_t)=E(y_{t+1}|\,\theta_t)=\theta_t + \mu [y^T-E(\theta_t)]$.
In this feedback-free scenario volatility in the policy action will add to the volatility of forecast errors, but will not cause bias.

As indicated above, the statistical properties of the equilibrium forecast can easily be derived from Proposition~\ref{proposition: statistical properties} by substituting the equilibrium forecast slope $c^\dag$ for $c$ and the equilibrium forecast intercept $b^\dag$ for $b$. Corollary \ref{corollary: equil bias MZ} presents the resulting formulas. 

\begin{corollary}\label{corollary: equil bias MZ}
Suppose that $\tau^2\le \frac{1}{4}$, and $\mu\neq \frac{1}{2}+\frac{\sqrt{1-4\tau^2}}{2}$ (so an equilibrium exists). Then: 
\begin{enumerate}
\item [(a)]  The conditional bias of the equilibrium forecast is given by
\begin{eqnarray}\label{eqn: equil bias corollary 2}
E(y_{t+1}-f^\dag_{t}|\,\theta_t)=\frac{1-\sqrt{1-4\tau^2}}{2} (\theta_t-y^T).
\end{eqnarray}
\item [(b)] The MZ regression function of the equilibrium forecast is given by
\begin{eqnarray*}
E(y_{t+1}|f^\dag_t)=\frac{\tau^2}{\tau^2- (1-\mu) \left(\frac{1}{2}+\frac{\sqrt{1-4\cdot \tau^2}}{2}\right)} y^T+ \frac{ (1-\mu) \left(\frac{1}{2}+\frac{\sqrt{1-4\cdot \tau^2}}{2}\right) }{ (1-\mu) \left(\frac{1}{2}+\frac{\sqrt{1-4\cdot \tau^2}}{2}\right) -\tau^2} f^\dag_t.
\end{eqnarray*}
\end{enumerate}

\end{corollary}

Corollary \ref{corollary: equil bias MZ} provides several further insights into forecaster behaviour and forecast properties in the presence of feedback.
First, part (a) of Corollary \ref{corollary: equil bias MZ} shows that our model is capable of generating sign-changing bias, which was the first motivating empirical pattern in the introduction. Specifically, by equation (\ref{eqn: equil bias corollary 2}), the sign of the equilibrium bias agrees with the sign of $\theta_t-y^T$ given that the multiplier in front of this term is always positive. If the state is higher than the target, then the expected equilibrium forecast error is positive, meaning that the forecaster tends to underpredict; i.e., he shrinks the forecast toward the target relative to the expected outcome. Conversely, if the state is below the target, then the expected forecast error is negative, implying that the forecaster tends to overpredict, again shrinking toward the target. In sum, the prediction of the model is that the optimal bias-variance tradeoff causes the equilibrium forecast to ``gravitate'' toward the target relative to the expected outcome. 

\begin{figure}[!t]
\begin{center}
\includegraphics[scale=.7]{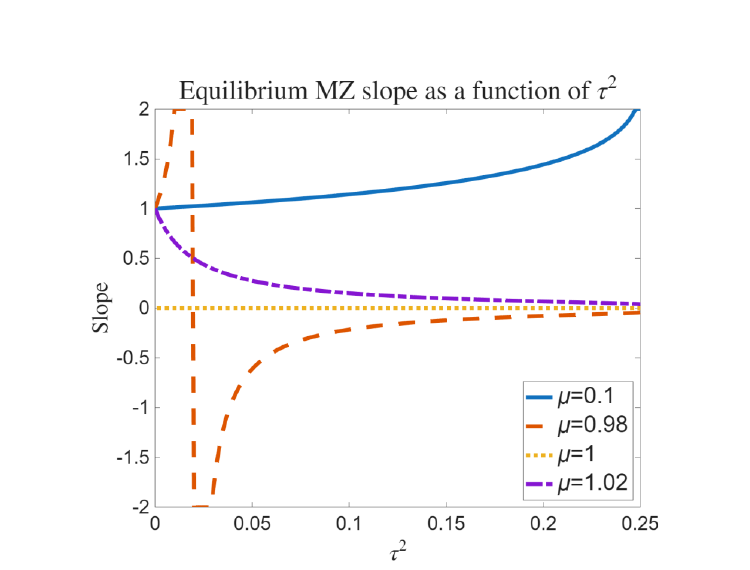}
\caption{The MZ slope of the equilibrium forecast as a function of $\tau^2$ for various values of $\mu$ (functions truncated at 2 and $-2$).}
\label{fig: equil MZ slopes tau2}
\end{center}
\end{figure}

Second, part (b) of Corollary \ref{corollary: equil bias MZ} shows that the equilibrium MZ intercept and slope are highly nonlinear functions of $\mu$ and $\tau^2$. Figure~\ref{fig: equil MZ slopes tau2} plots the slope as a function of $\tau^2$ for various fixed values of $\mu$. It is clearly seen that the slope can take on a wide range of different values for different configurations of $\mu$ and $\tau^2$; it can be large and positive, positive and close to zero, or even negative. Thus, the model is capable of accommodating the second motivating pattern from the introduction. 

Third, the broad interpretation of the parameter $\mu$ is that it measures the aggressiveness with which the DM pursues her target on average, while $\tau^2$ measures how predictable or consistent the DM is with regard to her policy choice. A value of $\mu$ close to one indicates that the DM takes whatever action is necessary to achieve her target, i.e., to close the gap between $\theta_t$ and $y^T$. A small value of $\mu$, on the other hand, means that the DM only undertakes small corrections on average. As Figure~\ref{fig: equil MZ slopes tau2} shows, if the DM switches from a cautious stance (e.g., $\mu$ around 0.1), to an aggressive stance (e.g., $\mu$ around 1), then we  expect to see a flattening of the equilibrium MZ slope provided that uncertainty is above a certain threshold (say, $\tau^2\ge 0.05)$. The same figure also shows that when $\mu$ is sufficiently close to 1, and there is already a moderate level of uncertainty, then further increases in $\tau^2$ attenuate the equilibrium MZ slope toward zero.

Fourth, when $\mu$ is precisely one, the MZ slope of the equilibrium forecast is zero, and the MZ intercept is $y^T$ for any value of $\tau^2$. This is because there is full adjustment toward the target on average, i.e., the average value of $y_{t+1}$ will be equal to $y^T$ for any value of the forecast, or more formally, $E(y_{t+1}|f^\dag_t)=y^T$ for all $f^\dag_t$. However, this result can be rather fragile for low values of $\tau^2$; in this case even a small deviation of $\mu$ from unity can induce a large positive or negative MZ slope (see Figure~\ref{fig: equil MZ slopes tau2}).

\section{Conclusion}\label{sec: conclusion}

We present and analyse a strategic model of forecasting in the presence of policy feedback. With the help of the model, we attempt to clarify the consequences of such feedback for the properties of observed forecasts. While these  environments are empirically relevant, to the best of our knowledge ours is the first formal treatment of the statistical properties of forecasts in these settings. Though our model is abstract, it opens the door for potential empirical investigations.

Our results show that when feedback effects are present, and there is some degree of uncertainty about the policy response to the forecast, the canonical properties of mean-square optimal forecasts break down. In particular, forecasts in feedback-prone environments are biased, and forecast errors are predictable given the information possessed by the forecaster. Furthermore, the slope of the Mincer-Zarnowitz regression can take on a wide range of values (including negative ones) depending on the average strength of the DM's reaction to the forecast and the uncertainty surrounding this reaction. These properties are in stark contrast to those of mean-squared optimal forecasts under no feedback. 

An important implication of our results is that the presence of feedback can confound the interpretation of forecast rationality tests as well as loss function estimation exercises.

\bibliographystyle{econometrica}
\bibliography{MS20250578-references}

\newpage

\appendix

\numberwithin{lemma}{section}
\numberwithin{theorem}{section}
\numberwithin{table}{section}
\numberwithin{equation}{section}
\numberwithin{corollary}{section}

\newcounter{asscountapp}
\numberwithin{asscountapp}{section}
\newenvironment{assumptionapp}
{\refstepcounter{asscountapp}\bigskip\noindent\textsc{Assumption~\Alph{section}.\arabic{asscountapp}}}
{\smallskip}

\pagestyle{plain}
\setcounter{page}{1}
\setcounter{footnote}{0}

\section*{Online Appendix}
{\bf ``Forecasting with Feedback'' by Lieli and Nieto-Barthaburu}\\
\smallskip

\noindent Throughout the appendix, we refer to equations or formulas in the main text as a single number in parenthesis; e.g., equation (2). We refer to equations in the appendix as (section.number); e.g., (B.2).  

Section~\ref{appsec: DM microfound} provides a microfoundation for the DM's reaction function (\ref{eqn: R's optimal action}). Section~\ref{appsec: equil select} presents the alternative linear equilibrium in the model and explains why we rule it out. Section~\ref{appsec: proofs} contains the proofs of all the formally stated results in the paper, including those in the Online Appendix.

\section{The DM's reaction function as the solution to an optimal control problem}\label{appsec: DM microfound}

Suppose that DM's loss from the outcome $y_{t+1}$ not hitting the target $y^T$ is a quadratic function of the control error $y_{t+1}-y^T$. In addition, the DM faces a quadratic adjustment cost from setting the policy variable to a nonzero value. Hence, the DM's total ex-post loss from her control problem has the form
\begin{eqnarray*}\label{eqn: DM's loss}
(y_{t+1}-y^T)^2+w_t\cdot a_t^2,
\end{eqnarray*}
where $w_t\in(-1,\infty)$ is the DM's adjustment cost parameter at time $t$. A high positive value of $w_t$ dissuades the DM from taking extreme action to hit the target. Conversely, negative values of the cost parameter correspond to a DM prone to overreaction. We make the same assumptions about $w_t$ as we do about $x_t$ in the main text. That is, the value of $w_t$ is drawn by ``nature'' from a known distribution, independently of $(\theta_t,\epsilon_{t+1})$, and the realised value of $w_t$ is the DM's private information. 

The DM's goal is to choose the action $a_t$ to minimise expected loss conditional on the information conveyed by the forecast. That is, for any given value of the forecast $f_t$, the DM chooses an action $a_t$ to solve
\begin{eqnarray*}\label{eqn: R's expected payoff}
\min_{a_t} \Big\{E\big[(y_{t+1}-y^T)^2\big|\,f_t\big]+w_t\cdot a_t^2\Big\}=\min_{a_t}\Big\{E\big[(\theta_t+a_t+\epsilon_{t+1}-y^T)^2\big|\,f_t\big]+w_t\cdot a_t^2\Big\},
\end{eqnarray*}
where the expectation is with respect to the distribution of $(\theta_t,\epsilon_{t+1})$ given $f_t$. Lemma~\ref{lemma: optimal action} below shows that the ``Taylor rule'' type of reaction function posited in equation (\ref{eqn: R's optimal action}) solves this problem, with $x_t=1/(1+w_t)$.
\begin{lemma}\label{lemma: optimal action} The optimal action of the DM is given by
\begin{eqnarray*}
a^*(f_t)=\frac{1}{1+w_t} \big[y^T-E(\theta_t|f_t)\big]:=x_t\big[y^T-E(\theta_t|f_t)\big],
\end{eqnarray*}
where $x_t=1/(1+w_t)>0$.
\end{lemma}

Thus, for a DM who faces the proposed tradeoff the reaction function (\ref{eqn: R's optimal action}) is a rational response to information revealed by the forecast. The proof of Lemma~\ref{lemma: optimal action} is given in Appendix \ref{appsec: proofs}.

\section{Equilibrium selection}\label{appsec: equil select} 

Here we provide an extension of Corollary \ref{corollary: equilibrium slopes}, which specifies both linear equilibria of the model.  

\begin{corollary}\label{app:corollary:equilibrium slopes}
For $\tau^2\le \frac{1}{4}$ there exist two linear PBE with equilibrium forecasts $f_i^\dag(\theta_t)=b^\dag_i+c^\dag_i\cdot\theta_t$, $i=1,2$.
\begin{enumerate}[(1)]
\item The slopes are given by:
\begin{align*}
c^\dag_1 =\frac{1}{2}-\mu+\frac{\sqrt{1-4\cdot \tau^2}}{2},
\end{align*}
and
\begin{align*}
c^\dag_2 =\frac{1}{2}-\mu-\frac{\sqrt{1-4\cdot \tau^2}}{2},
\end{align*}
provided that $c^\dag_1\neq 0$ and $c^\dag_2\neq 0$.

\item The intercepts are given by:
\begin{align*}
b_i^\dag = (1-c^\dag_i) y^T,\qquad i=1,2.
\end{align*}
\end{enumerate}
\end{corollary}

In order to use the model for interpreting observed forecast patterns, we need to take a stance on which equilibrium provides more plausible predictions empirically. Before we proceed to justify our choice, we first derive the bias and loss under each equilibrium.

\begin{corollary}\label{app: corollary: bias-Pareto preference}
For the equilibria specified in Corollary \ref{app:corollary:equilibrium slopes}:
\begin{enumerate}[(1)]
    \item The equilibrium conditional bias is
\begin{align*}
bias^\dag_1=E[y_{t+1}-f_1^\dag(\theta_t)\big|\, \theta_t] =  \frac{1-\sqrt{1-4\tau^2}}{2}(\theta_t-y^T),
\end{align*}
and
\begin{align*}
bias^\dag_2 = E[y_{t+1}-f_2^\dag(\theta_t)\big|\, \theta_t]=\frac{1+\sqrt{1-4\tau^2}}{2}(\theta_t-y^T).
\end{align*}

\item The equilibrium conditional loss is
\begin{align*}
loss^\dag_1=E[(y_{t+1}-f_1^\dag)^2\big|\,\theta_t] = \frac{1-\sqrt{1-4\tau^2}}{2}(\theta_t-y^T)^2+\sigma^2,
\end{align*}
and
\begin{align}
loss^\dag_2=E[(y_{t+1}-f_2^\dag)^2\big|\,\theta_t] = \frac{1+\sqrt{1-4\tau^2}}{2}(\theta_t-y^T)^2+\sigma^2.
\end{align}

\end{enumerate}
\end{corollary}

We argue that the more reasonable equilibrium in our model is the first one, i.e., with forecast slope $c_1^\dag$ and intercept $b_1^\dag$, for the following reasons.
\begin{enumerate}
\item The first equilibrium is Pareto preferred to the second one. To see this, first note that the DM learns the state in both equilibria; hence, the DM's action, and therefore her welfare, is the same in both situations. Next, part (2) of Corollary \ref{app: corollary: bias-Pareto preference} gives an explicit expression for the equilibrium loss of the forecaster in each case. It is immediate from these expressions that the loss is lower in the first equilibrium. Hence, the players are jointly better off in that situation (the DM is equally well off, and the forecaster is strictly better off).

\item Next, our chosen equilibrium has a forecast that is less biased (in absolute value), as part (1) of Corollary \ref{app: corollary: bias-Pareto preference} reveals. Indeed, the proof of the corollary (provided in Appendix \ref{appsec: proofs}) shows that the variance term in the forecaster's MSE decomposition (given in equation (\ref{eq: mse decomposition}) in the main text) takes the same value in both equilibria. Therefore, the ranking of equilibrium losses agrees with the ranking of absolute biases; i.e., lower equilibrium bias is equivalent to lower equilibrium forecaster loss.

\item Finally, consider the limit of the two equilibria as uncertainty about $x_t$ vanishes, i.e., as $\tau^2\rar 0$. As the DM learns the true value of $\theta_t$ from the forecast in both cases, in the limit the DM's optimal action is $a^*_t=\mu(y^T-\theta_t)$, and the resulting outcome is $y_{t+1}=\theta_t+\mu(y^T-\theta_t)+\epsilon_{t+1}$. The ``natural'' optimal forecast of this outcome is then $\theta_t+\mu(y^T-\theta_t)$, which is indeed the limit of $f_{1}^\dag(\theta_t)$ as $\tau^2\rar 0$. But how can $f^\dag_{2}(\theta_t)$ also be an equilibrium forecast? It can be verified that $\lim_{\tau^2\rar 0} f^\dag_2(\theta_t)=y^T+\mu(y^T-\theta_t)$. The corresponding reaction function (\ref{eqn: R's optimal action with exp}) is then given by $a^*(f)=f-y^T$. But if the DM uses this rule, it is immediately seen that the forecaster's objective (\ref{eqn: fc prob general}) becomes completely independent of the forecast. So, he might as well report $\lim f^\dag_{2}(\theta_t)$ which, of course, justifies the DM's decision rule. This is an odd case, \label{rem: equil selection} in the sense that forecaster behaviour in equilibrium rests on the fact that the DM chooses an action that makes the forecaster completely indifferent among all forecasts.\footnote{When $\tau^2$ is exactly zero, there exist many more equilibria that are supported by the DM making the forecaster indifferent, including a family of \emph{pooling} equilibria where the forecast is completely uninformative, i.e., a constant function of $\theta$. As we expect that in practice there will be at least some uncertainty about the DM's reaction, we do not deem these equilibria relevant. Furthermore, this sort of forecasting practice does not match observed behaviour.}  

\end{enumerate}
For these reasons, we choose the first PBE, with slope (\ref{eqn: equilibrium slope}) and intercept (\ref{eqn: equilibrium intercept}), as our preferred prediction about behaviour in the forecasting game. 

\section{Proofs}\label{appsec: proofs}

For notational simplicity, in the proofs we drop the time ($t$) subscripts from the variables $y_{t+1}$, $f_t$, $\theta_t$, $x_t$ and $\epsilon_{t+1}$. This should not cause any confusion. 

\subsection*{Proof of Proposition \ref{prop: non-strat opt forecast}}

To see the decomposition in equation (\ref{eq: mse decomposition}), write 
\begin{eqnarray*}
&&E[(y-f)^2\big|\,\theta] =  E\Big\{\big[y-E(y|\theta)+E(y|\theta)-f\big]^2\big|\,\theta\Big\}\\
&&=E\Big\{\big[y-E(y|\theta)\big]^2\big|\,\theta\Big\}+E\Big\{[E(y|\theta)-f]^2\big|\,\theta\Big\}+2E\Big\{[y-E(y|\theta)]\cdot [E(y|\theta)-f]\big|\,\theta\Big\}.
\end{eqnarray*}
Since $E(y|\theta)$ is a function of $\theta$ alone, we have that $E\{[E(y|\theta)-f]^2\big|\,\theta\}=[E(y|\theta)-f]^2$ and 
\[
E\Big\{[y-E(y|\theta)]\cdot [E(y|\theta)-f]\big|\,\theta\Big\}=[E(y|\theta)-f]\big\{E(y|\theta)-E(y|\theta)\big\}=0,
\]
which gives the first equality in (\ref{eq: mse decomposition}). The last equality in (\ref{eq: mse decomposition}) follows because
\[
Var(y|\theta)=Var[\theta + a^*(f)+\epsilon|\,\theta]=Var[a^*(f)+\epsilon|\,\theta]=Var[a^*(f)|\,\theta]+\sigma^2,
\]
where in the last step we use the fact that $\epsilon$ and $a^*(f)$ are independent conditional on $\theta$.\footnote{In the text we assume that $\epsilon$ is independent of $\theta$ (after equation (\ref{eqn: outcome equation})) and that $(\theta,\epsilon)$ is independent of $x$ (in Section~\ref{subsec: DM's problem}). This implies mutual independence of $x$, $\theta$ and $\epsilon$, which, in turn, implies that $x$ and $\epsilon$ are independent conditional on $\theta$.} Equations (\ref{eqn: var expansion}) and (\ref{eqn: bias expansion}) in the main text can then be derived as 
\begin{eqnarray*}
    &&Var(y|\,\theta)=Var\big[a^*(f)\big|\,\theta \big]+\sigma^2=Var\left[x\left(y^T-\frac{f}{c}+\frac{b}{c}\right) \Big|\,\theta \right] +\sigma^2 \nonumber\\
    &&\quad=Var(x|\,\theta) \left(y^T-\frac{f}{c}+\frac{b}{c}\right)^2 + \sigma^2=\tau^2 \left(y^T-\frac{f}{c}+\frac{b}{c}\right)^2 + \sigma^2,
\end{eqnarray*}
and
\begin{eqnarray*}
    &&bias^2(f|\,\theta)=\big\{\theta+E[a^*(f)|\theta]-f\big\}^2 = \left\{\theta+E\left[x\left(y^T-\frac{f}{c}+\frac{b}{c}\right) \Big|\,\theta \right]-f\right\}^2\nonumber\\ 
    &&\quad=\left\{\theta+E(x|\,\theta)\left(y^T-\frac{f}{c}+\frac{b}{c}\right) -f\right\}^2=\left\{\theta+ \mu \Big(y^T+\frac{b}{c}\Big)-\Big(\frac{\mu}{c}+1\Big)f\right\}^2,
\end{eqnarray*}
where the independence of $x$ and $\theta$ is used in multiple steps. Thus, the forecaster's conditional MSE is given by
\begin{eqnarray}
    Var(y|\theta)+bias^2(f|\theta)
    &=&\tau^2\left(y^T-\frac{f}{c}+\frac{b}{c}\right)^2+\sigma^2+\left[\theta+\mu\left(y^T+\frac{b}{c}\right)-\left(\frac{\mu}{c}+1\right)f\right]^2. \label{app: MSE given theta}
\end{eqnarray}
Setting the partial derivative of (\ref{app: MSE given theta}) w.r.t.\ $f$ equal to zero results in the FOC:
\begin{equation*}
    \frac{\tau^2}{c}\left(y^T-\frac{f}{c}+\frac{b}{c}\right)+\left(\frac{\mu}{c}+1\right)\left[\theta+\mu\left(y^T+\frac{b}{c}\right)-\left(\frac{\mu}{c}+1\right)f\right]=0;
\end{equation*}
solving this linear equation for $f$ yields $f^*(\theta)=d^*+e^*\cdot\theta$, where the the intercept $d^*$ is given by expression (\ref{eqn: non-strat intercept}) and the slope $e^*$ is given by expression (\ref{eqn: non-strat slope}).\eprf 

\subsection*{Proof of Corollary \ref{corollary: equilibrium slopes}}

As explained in Remark 1 after Corollary \ref{corollary: equilibrium slopes}, to find the equilibrium forecast slope we set
\begin{equation*}
    c^\dag=\frac{c^\dag(\mu+c^\dag)}{\tau^2+(\mu+c^\dag)^2}.
\end{equation*}
Dropping the dagger superscript and rearranging yields the following quadratic equation in $c$:
\[
c^2+(2\mu-1)c+\mu(\mu-1)+\tau^2=0.
\]
The quadratic formula then gives:
\[
c_{1,2}=\frac{-(2\mu-1)\pm\sqrt{(2\mu-1)^2-4(\mu(\mu-1)+\tau^2)} }{2}=\frac{1}{2}-\mu \pm \frac{1}{2}\sqrt{1-4\tau^2}.
\]
The value $c_1$ is the equilibrium slope stated in Corollary~\ref{corollary: equilibrium slopes} (with the `$+$' sign) and the value $c_2$ is the alternative equilibrium slope presented in Corollary~\ref{app:corollary:equilibrium slopes} (with the `$-$' sign). Here, we will continue to work with $c^\dag=c_1$. 

To obtain the equilibrium intercept, we first substitute the common value $b^\dag$ for $d^*$ and $b$ in equation (\ref{eqn: non-strat intercept}) to obtain
\[
b^\dag=\frac{\tau^2+\mu(\mu+c^\dag)}{\tau^2+(\mu+c^\dag)^2}\left(c^\dag y^T+b^\dag\right),
\]
where we also set $c=c^\dag$ to fully impose equilibrium conditions. To facilitate further calculations we define $$A:=\frac{1+\sqrt{1-4\tau^2}}{2}.$$ It is straightforward to check that $c^\dag=A-\mu$ and 
$A^2+\tau^2=A$ or, equivalently, $\tau^2=A(1-A)$.

Next, dropping the dagger superscripts, and using the properties of $A$ above we obtain
\begin{align*}    
b=&\frac{\tau^2+\mu(\mu+c)}{\tau^2+(\mu+c)^2}\left(cy^T+b\right)=\frac{A(1-A)+\mu A}{A}\left(cy^T+b\right)\\
=&(1-A+\mu) \left(cy^T+b\right)=(1-c)\left(cy^T+b\right).
\end{align*}
Solving for $b$ then yields
\[
b=(1-c)y^T
\]
as stated in (\ref{eqn: equilibrium intercept}). \eprf

\subsection*{Proof of Proposition~\ref{proposition: statistical properties}}

Starting from the bias formula in equation (\ref{eqn: bias expansion}) and substituting $f^*=d^*+e^*\cdot\theta$ gives
\begin{eqnarray*}
   E(y|\,\theta)-f^*&=&\theta+\mu\left(y^T+\frac{b}{c}\right)-\left(\frac{\mu}{c}+1\right)(d^*+e^*\cdot\theta)\\
   &=& \mu\left(y^T+\frac{b}{c}\right)-\left(\frac{\mu}{c}+1\right)d^*+\left[1-e^*\left(\frac{\mu}{c}+1\right)\right]\theta.
\end{eqnarray*}
Substituting expression (\ref{eqn: non-strat slope}) for $e^*$ and expression (\ref{eqn: non-strat intercept}) for $d^*$ yields
\begin{eqnarray*}
   &&E(y|\,\theta)-f^*\\
   &&=\mu\left(y^T+\frac{b}{c}\right)
   -\left(\frac{\mu}{c}+1\right)c\frac{\tau^2+\mu(\mu+c)}{\tau^2+(\mu+c)^2}\left(y^T+\frac{b}{c}\right)
   +\left[1-\left(\frac{\mu}{c}+1\right)\frac{c(\mu+c)}{\tau^2+(\mu+c)^2}\right]\theta\\
   &&=\left(y^T+\frac{b}{c}\right)\frac{-c\tau^2}{\tau^2+(\mu+c)^2}+\frac{\tau^2}{\tau^2+(\mu+c)^2}\theta\\
   &&=\frac{\tau^2}{\tau^2+(\mu+c)^2}(\theta-cy^T-b),
\end{eqnarray*}
which is the result stated in Proposition~\ref{proposition: statistical properties} part (a). 

Turning to the Mincer-Zarnowitz (MZ) regression, we write the outcome as
\begin{equation}\label{app: outcome detailed}
    y=\theta+x\left(y^T-\frac{f^*}{c}+\frac{b}{c}\right)+\epsilon.
\end{equation}
One can take the expectation of (\ref{app: outcome detailed}) conditional on $f^*=f^*(\theta)$, to obtain
\begin{eqnarray}\label{app: MZ}
E(y|f^*)=E(\theta|f^*)+\mu\left(y^T-\frac{f^*}{c}+\frac{b}{c}\right),
\end{eqnarray}
where we used the fact that $f(\theta)$ is independent of $x$ and $\epsilon$ (because is $\theta$ is independent of $x$ and $\epsilon$). Since the optimal forecast is $f^*=d^*+e^*\theta$, it follows that $E(\theta|f^*)=f^*/e^*-d^*/e^*$. Substituting into (\ref{app: MZ}) gives
\begin{eqnarray*}
E(y|f^*)&=&\frac{f^*}{e^*}-\frac{d^*}{e^*}+\mu\left(y^T-\frac{f^*}{c}+\frac{b}{c}\right)=\left(\frac{1}{e^*}
-\frac{\mu}{c}\right)f^*+\mu \left(y^T+\frac{b}{c}\right)-\frac{d^*}{e^*}.
\end{eqnarray*}
Further substituting expression (\ref{eqn: non-strat slope}) for $e^*$ and expression (\ref{eqn: non-strat intercept}) for $d^*$ yields
\begin{eqnarray*}
E(y|f^*)&=&\left(\frac{\tau^2+(\mu+c)^2}{c(\mu+c)}-\frac{\mu}{c}\right)f^*+\mu \left(y^T+\frac{b}{c}\right)-\frac{\left(y^T+\frac{b}{c}\right)(\tau^2+\mu(\mu+c))}{\mu+c}\\
&=&\frac{\tau^2+c(\mu+c)}{c(\mu+c)}f^*+\left(y^T+\frac{b}{c}\right)\left(\mu-\frac{\tau^2+\mu(\mu+c)}{\mu+c}\right)\\
&=&-\left(\frac{\tau^2}{\mu+c}\right)\left(y^T+\frac{b}{c}\right)+\frac{\tau^2+c(\mu+c)}{c(\mu+c)}f^*,
\end{eqnarray*}
as stated in equation (\ref{eq: MZ}). \eprf

\subsection*{Proof of Corollary~\ref{corollary: equil bias MZ}}
Equations (\ref{eqn: equilibrium slope}) and (\ref{eqn: equilibrium intercept}) specify the equilibrium forecast slope $c^\dag$ and intercept $b^\dag$ as functions of $\mu$, $\tau^2$ and $y^T$. We will substitute these expressions in place of $c$ and $b$ in the optimal bias formula (\ref{eq: bias}) and the Mincer-Zarnowitz regression formula (\ref{eq: MZ}). As in the proof of Corollary \ref{corollary: equilibrium slopes} above, we again let $$A:=\frac{1+\sqrt{1-4\tau^2}}{2}.$$ Recall that with this definition, $c^\dag=A-\mu$ and $A^2+\tau^2=A$ or, equivalently, $\tau^2=A(1-A)$. These simple algebraic properties further imply $Ac^\dag=(1-\mu)A-\tau^2$.

First, we will derive the expression for the conditional bias. Applying the optimal bias formula (\ref{eq: bias}) to the equilibrium forecast gives 
\[
E(y|\theta)-f^\dag=\frac{\tau^2}{\tau^2+(\mu+c^\dag)^2}(\theta-c^\dag y^T-b^\dag).
\]
Using the fact that the equilibrium slope is $c^\dag=A-\mu$, and the properties of $A$, we obtain
\begin{eqnarray*}
    \frac{\tau^2}{\tau^2+(\mu+c^\dag)^2}=\frac{\tau^2}{\tau^2+A^2}=\frac{A(1-A)}{A}=1-A=\frac{1-\sqrt{1-4\tau^2}}{2}.
\end{eqnarray*}
Next, using equation (\ref{eqn: equilibrium intercept}),
\begin{eqnarray*}
    c^\dag y^T+b^\dag=c^\dag y^T+(1-c^\dag)y^T=y^T,
\end{eqnarray*}
and so 
\[
E(y|\theta)-f^\dag=\frac{\tau^2}{\tau^2+(\mu+c^\dag)^2}(\theta-c^\dag y^T-b^\dag)=\frac{1-\sqrt{1-4\tau^2}}{2}(\theta-y^T)
\]
as stated in part (a) of Corollary \ref{corollary: equil bias MZ}.

Turning to the slope of the equilibrium MZ regression function, we substitute the equilibrium forecast slope $c^\dag$ for $c$ in equation (\ref{eq: MZ}). This gives
\[
\frac{\tau^2+c^\dag(\mu+c^\dag)}{c^\dag(\mu+c^\dag)}=\frac{\tau^2+c^\dag A}{c^\dag A}=\frac{(1-\mu)A}{(1-\mu)A-\tau^2},
\]
which is precisely the expression stated in part (b) of Corollary \ref{corollary: equil bias MZ}. Similar calculations lead to the formula for the MZ intercept.\eprf

\subsection*{Proof of Lemma \ref{lemma: optimal action}} 

Let $a$ be a fixed action. As $\theta$ and $\epsilon$ are assumed to be independent, it is easy to see that 
\[
E\big[(\theta+a+\epsilon-y^T)^2\big|\,\theta\big]=(\theta+a-y^T)^2+\sigma^2.
\]
As $f$ is a function of $\theta$, the law of iterated expectations yields
\[
E\big[(\theta+a+\epsilon-y^T)^2\big|\,f\big]=E[(\theta+a-y^T)^2|\,f]+\sigma^2.
\]
Thus, we can rewrite the DM's objective as
\begin{eqnarray*}
&&E\big[(\theta+a-y^T)^2\big|\,f\big]+\sigma^2+w\cdot a^2=E\big[(\theta-y^T)^2\big|\,f\big]+2E(\theta-y^T|\,f)a+\sigma^2+a^2+w\cdot a^2.
\end{eqnarray*}
Taking the partial derivative with respect to $a$ and setting it to zero gives
\[
2[E(\theta|f)-y^T]+2(1+w)a=0.
\]
Solving for $a$ gives the formula for the optimal action stated in Lemma~\ref{lemma: optimal action}. The second order condition for a minimum is clearly satisfied as long as $w>-1$.\eprf

\subsection*{Proof of Corollary \ref{app:corollary:equilibrium slopes}} The equilibrium with forecast function $f^\dag_1(\theta)=b_1^\dag+c_1^\dag\theta$ was already shown to exist in the proof of Corollary~\ref{corollary: equilibrium slopes}. The same proof also contains the formula for $c_2^\dag$. The result $b_2^\dag=(1-c_2^\dag)y^T$ follows from the fact that if we redefine $A:=\frac{1}{2}(1-\sqrt{1-4\tau^2})$ in the proof of Corollary~\ref{corollary: equilibrium slopes}, then $c_2^\dag=A-\mu$ and $A$ retains all additional algebraic properties stated there.\eprf

\subsection*{Proof of Corollary \ref{app: corollary: bias-Pareto preference}}

The two equilibrium forecasts are $f^\dag_i=b^\dag_i+c^\dag_i\theta$, $i=1,2$. Let
$$A_1:=\frac{1+\sqrt{1-4\tau^2}}{2}  \qquad \text{and} \qquad A_2:=\frac{1-\sqrt{1-4\tau^2}}{2},$$ 
so that $c^\dag_i=A_i-\mu$ and $A_i^2+\tau^2=A_i$ and $A_ic^\dag_i=(1-\mu)A_i-\tau^2$ for $i=1,2$. 
The formula for $bias_1^\dag$ is derived in the proof of Corollary~\ref{corollary: equil bias MZ} (note that $A=A_1$). Replacing $A$ with $A_2$ in the same derivations gives the formula for $bias_2^\dag$. 

To compute the equilibrium losses, first note that by equation (\ref{eqn: var expansion})
\[
Var^\dag(y|\,\theta)=\tau^2 \left(y^T-\theta\right)^2 + \sigma^2
\]
in both equilibria because 
$$\frac{f^\dag_i}{c^\dag_i}+\frac{b^\dag_i}{c^\dag_i}=\theta$$ for $i=1,2$. Next,
\begin{eqnarray*}
    loss^\dag_1&=&Var^\dag(y|\,\theta)+(bias^\dag_1)^2\\
    &=&\tau^2 \left(y^T-\theta\right)^2 + \sigma^2 + A_2^2(y^T-\theta)^2= (\tau^2 + A_2^2)(y^T-\theta)^2 + \sigma^2=A_2(y^T-\theta)^2 + \sigma^2, 
\end{eqnarray*}
and analogously for $loss^\dag_2$.\eprf

\end{document}